\documentclass[]{jfm}

\usepackage{graphicx}
\usepackage{epstopdf,epsfig}
\usepackage{enumerate}
\usepackage{epstopdf}
\usepackage{epsfig}
\usepackage{subfigure}
\usepackage{tikz}
\usepackage{newtxtext}
\usepackage{newtxmath}
\usepackage{natbib}
\usepackage{hyperref}
\hypersetup{
    colorlinks = true,
    urlcolor   = blue,
    citecolor  = black,
}

\newcommand{\RomanNumeralCaps}[1]
\linenumbers

\title{Larger wavelengths suit the hydrodynamics of carangiform swimmers}

\author{Muhammad Saif Ullah Khalid\aff{1,2,3},
  Junshi Wang\aff{4},
  Imran Akhtar\aff{5},
  Arman Hemmati\aff{3},
  Haibo Dong\aff{4},
 \and Moubin Liu\aff{1,2}
 \corresp{\email{mbliu@pku.edu.cn}}}

\affiliation{\aff{1}Institute of Ocean Research, Peking University, Beijing, People's Republic of China
\aff{2}Key State Laboratory of Complex Flows and Turbulence, Department of Mechanics, Peking University, Beijing, People's Republic of China
\aff{3} Department of Mechanical Engineering, University of Alberta, Edmonton, AB, Canada T6G 1H9
\aff{4}Department of Mechanical and Aerospace Engineering, University of Virginia, Charlottesville, Virginia 22904, USA
\aff{5}Department of Mechanical Engineering, NUST College of Electrical \& Mechanical Engineering, National University of Sciences and Technology, Rawalpindi, Pakistan}

\begin{document}
\maketitle

\begin{abstract}
The wavelength of undulatory kinematics of fish is an important parameter to determine their hydrodynamic performance. This study focuses on numerical examination of this feature by reconstructing the real physiological model and kinematics of steadily swimmning Jack Fish. We perform three-dimensional numerical simulations for flows over these models composed of the trunk, and dorsal, anal, and caudal fins. Moreover, we prescribe the carangiform-like motion for its undulation for a range of wavelengths. Undulating with larger wavelengths improves the hydrodynamic performance of the carangiform swimmer in terms of better thrust production by the caudal fin, lower drag production on the trunk, and reduced power consumption by the trunk. This coincides with the formation of stronger posterior body vortices and leading-edge vortices with more circulation on the caudal fin. The real kinematics of Jack Fish surpasses the performance of those with prescribed motion owing to the flexibility of the caudal fin.

\end{abstract}


{\bf MSC Codes }  {\it(Optional)} Please enter your MSC Codes here

\section{Introduction}
\label{sec:intro}
Considering their gaits and kinematics, fish are classified on the basis of the wavelength ($\lambda$) with which they choose to undulate their bodies for propulsion. Most commonly known classes are anguilliform, subcarangiform, carangiform, and thunniform. As explained by \citet{Sfakiotakis1999} and \citet{Lauder2006a}, anguilliform swimmers undulate their bodies at wavelengths ($\lambda$) shorter than their body-lengths ($L$), that is $\lambda << L$. In case of subcarangiform and carangiform, $\lambda$ is observed to be either equal to or slightly larger than $L$, i.e., $\lambda/L \approx 1$. Thunniform swimmers primarily employ their caudal fins with almost no oscillations of their trunks and $\lambda/L >> 1$. 

Bio-inspired underwater robots can be designed using the fish physiology and kinematics to match their hydrodynamic performance \citep{Zhu2019, Fish2020, White2020}. To this end, it becomes imperative to investigate the link between fish morphological structures and their gaits. Previously, \citet{Liu2017} analyzed the effect of median fins on the thrust production capability of Jack Fish (\textit{Crevalle jack}). They concluded that the leading-edge vortex ($\mbox{LEV}$) on the caudal fin, which is the primary contributor to thrust production, on the caudal fin became stronger due to its interference with the vortices generated by the posterior body region. Later on, \citet{Han2020} investigated the role of the shape and motion of dorsal and anal fins of sunfish (\textit{Lepomis macrochirus}). When they enhanced the area of median fins with adjustments in the phase of their oscillatory motion, the thrust and efficiency of the caudal fin were simultaneously improved by $25.6\%$ and $29.2\%$, respectively. Recently, \citet{Wang2020} performed simulations for flow over finlets of yellowfin tuna (\textit{Thunnus albacares}) which showed that these small pitching fins could reduce drag by $21.5\%$ and power consumption by $20.8\%$. 

There remains a lack of knowledge regarding the role of wavelength of undulatory motion of fish in determining their thrust production capacity and power consumption, especially in relation to their real physiology. Addressing this knowledge gap becomes more significant when determining the required flexibility at robotic joints to design and operate bio-inspired vehicles. This also contributes greatly to our understanding of the evolution biology of fish, in which they have achieved such specific motion and flexibility for optimum swimming. Earlier, \citet{Borazjani2010} performed numerical simulations to examine how the hydrodynamic performance of mackerel (\textit{Scomber scombrus}), whcih is a carangiform swimmer, and lamprey, which is an anguilliform swimmer, would be affected if their respective kinematic patterns were exchanged. However, their study was limited due to the absence of mackarel's median fins in their physiological models. Also, the choice of $\lambda$ was restricted to only two values; $\lambda=0.642$ and $0.95$. Considering a foil as the representative cross-section of a fish, \citet{Khalid2020a} conducted numerical investigations to understand the effect of waveform on hydrodynamic performance parameters including thrust, power consumption, and efficiency. They found that anguilliform swimmers outperformed carangiform ones when they swim at $\lambda > L$. However, carangiform swimmers produced more thrust and efficiency for $\lambda < L$. These findings showed that natural swimmers might choose their $\lambda$'s restricted by other biological needs and not to enhance their swimming performance. In this work, we further build onto this study and perform high-fidelity simulations for real Jack Fish with both original and prescribed kinematics. Here, we explain why large wavelenghts ($\lambda^\star = \lambda / L \approx 1$) suit the hydrodynamics of Jack Fish representing carangiform swimmer.              

\section{Computational Methodology}
\label{sec:comp_method}

\subsection{Physiological Model and Kinematics of Crevalle Jack}
To reconstruct physiological structures of the trunk and median and caudal fins of Jack Fish and its kinematics, we employ the data recorded and reported previously by \citet{Liu2017}. Although the procedure to capture the fish motion and its geometry along with the statistical details has been covered in Ref. \citep{Liu2017}, we present its salient points here as well for the sake of completeness. The current model is of Crevalle Jack (Caranx hippos) which is classified as a carangiform swimmer. Out of total $\mbox{12}$ individuals of this class of fish with a mean total length $L=0.338\mbox{m}$ and swimming at $1\mbox{L/sec}$ to $4\mbox{L/sec}$. It is important to highlight that their body kinematics did not change much with the increasing swimming speed. The currently used kinematic data was adopted from an individual fish having $L=0.31\mbox{m}$ and swimming at $2\mbox{L/sec}$. The total height and width, normalized by $L$, of this fish are $0.286$ and $0.144$, respectively. The area, normalized by $L^2$, of the caudal fin is $0.023$ and its normalized length is $0.244$. The normalized height and length of the caudal fin are $0.315$ and $0.244$, respectively. 

\begin{figure}
  \centerline{\includegraphics[scale=0.1]{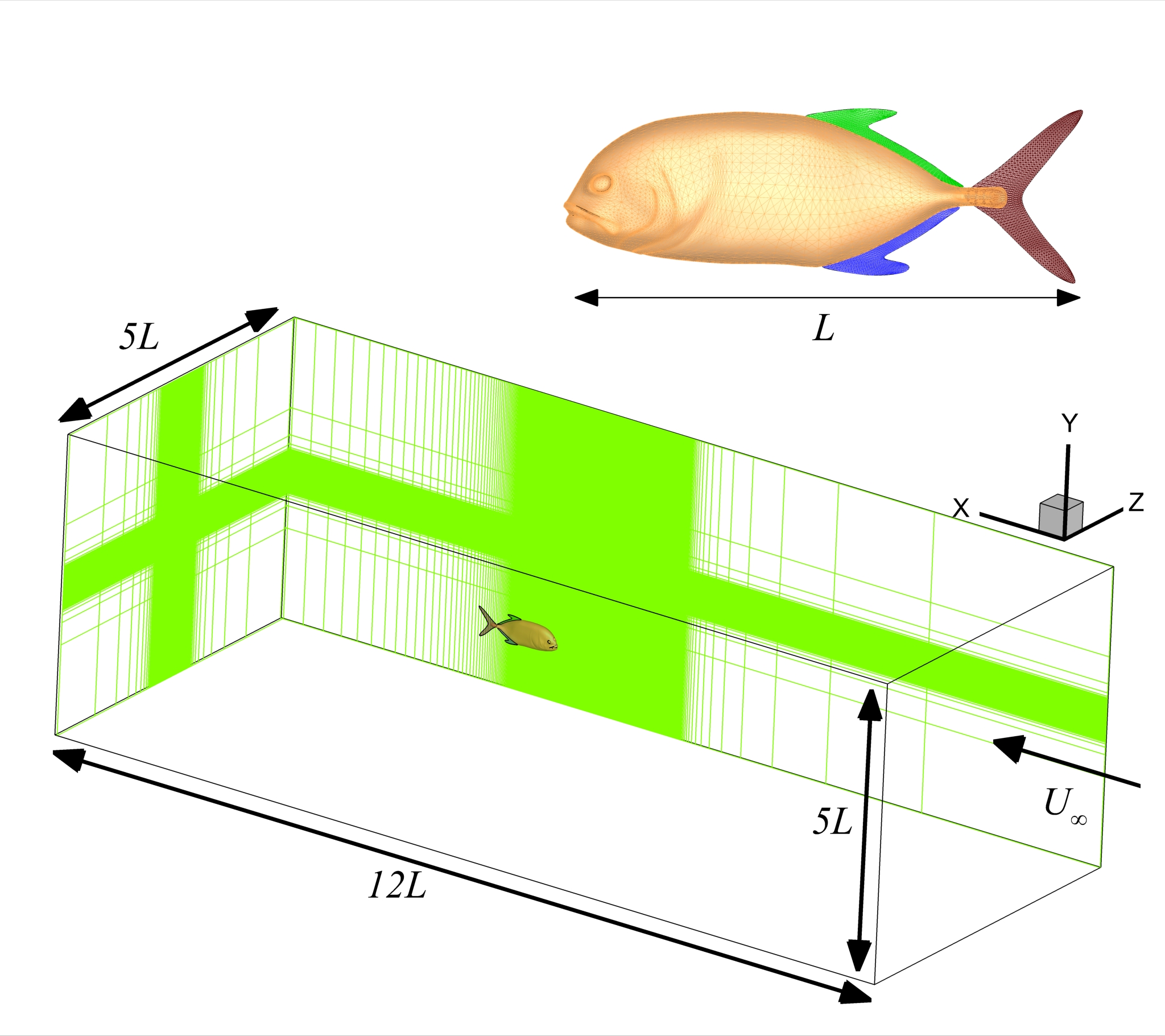}}
  \caption{Virtual tunnel for simulating flows over Jack Fish with its dimensions and the physilogical model of the fish covered with a mesh to indicate the marker points to track their motion}
\label{fig:flow_domain}
\end{figure}

In this study, we add median fins; the dorsal and anal fins, as well because these flexible membranous structures also contribute towards the propulsive functionality of a fish. Its trunk is modeled as a solid body with a closed surface and the dorsal, anal, and caudal fins are membranes with zero thickness. Each surface is, then, represented by triangular mesh where the main body is composed of 11358 nodes and 22712 elements. The surface of the caudal fin has 1369 nodes and 2560 elements. The dorsal and anal fins are composed of 885 and 895 elements with 1664 and 1680 elements, respectively (see Fig.~\ref{fig:flow_domain}). The measured original wavelength from the midline profiles is approximately 1.05L which is a characteristic of the carangiform swimming mode. The measured Strouhal number (\mbox{St}) for these recordings remain 0.30, where $\mbox{St}={2{A_\circ}{f}}/{U_\infty}$ with $f$ being the excitation/flapping frequency of the caudal fin, $A_\circ$ as the maximum one-sided oscillation amplitude of the caudal fin; also a measure of the wake-width, and $U_\infty$ as the free-stream velocity. 

In our present work, we perform three-dimensional ($\mbox{3D}$) numerical simulations for flows over the complete physiological model of Jack Fish with its real and prescribed kinematics with $0.65 \ge \lambda \le 1.25$. The carangiform amplitude profile is given by the following relation \citep{Khalid2016, Khalid2018, Khalid2020a}.

\begin{align}
A(\frac{x}{L}) &= 0.02 - 0.0825(\frac{x}{L}) + 0.1625(\frac{x}{L})^2; 0 < \frac{x}{L} < 1
\label{eq:carangiform}
\end{align}

Where $A(x/L)$  denotes the local amplitude at a given spatial position along the fish body; nondimensionalized by its total length ($L$). Here, The coefficients are calculated based on the data provided for a steadily swimming saithe fish; another carangiform swimmer \citep{Videler1993} where the local amplitudes are $A(0)=0.02$, $A(0.2)=0.01$, and $A(1.0)=0.10$. 

The undulatory kinematics takes the following form in both the cases.

\begin{align}
z(x/L,t) &= A(x/L)\sin[2\pi(x/\lambda - {f}{t}) + \phi]
\label{eq:prescribed_kinematics}
\end{align}

\begin{figure}
  \centerline{\includegraphics[scale=0.55]{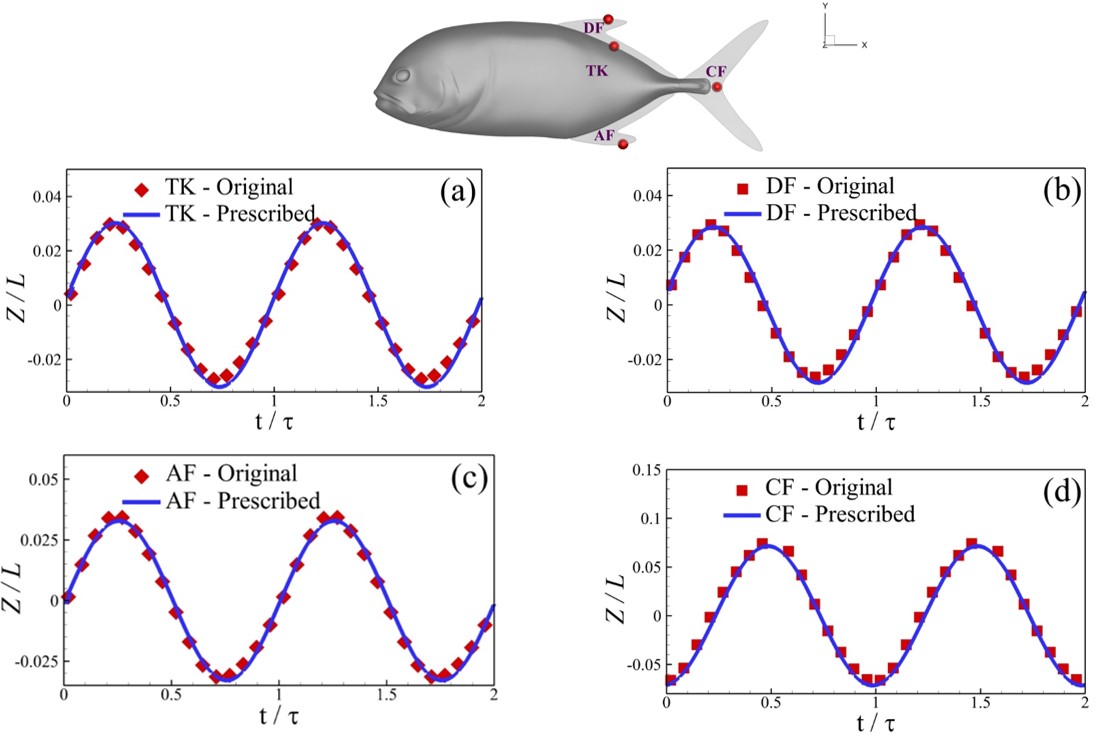}}
  \caption{Comparison of the real and prescribed kinematics ($\lambda^\star=1.05$ and $\phi=-5.5^\circ$) through temporal histories of displacements for four distinct points on the trunk and dorsal, anal, and caudal fins}
\label{fig:kin_comp}
\end{figure}

Here, the factor $2\pi/\lambda$ defines the wave-number ($k$) for the waveform of the kinematic profile along the swimmer’s body and $\phi$ denotes the phase of oscillation. In Fig.~\ref{fig:kin_comp}, we present the comparison of real and prescribed ($\lambda^\star=1.05$) kinematics with $\phi=-5.5^\circ$ for four different points on the trunk and anal, dorsal, and caudal fins. It is clear that Eq.~\ref{eq:prescribed_kinematics} for the prescribed motion mimics the real kinematics very well. 

\subsection{Numerical Solver}
We perform three dimensional ($\mbox{3D}$) numerical simulations for $\mbox{Re}=3000$ and $\mbox{St}=0.33$ using our in-house sharp-interface immersed boundary method based computational solver \citep{Mittal2008, Liu2017} that is suitable for both rigid and membranous body-structures. The reason behind the selection of this $\mbox{St}$ is that the real fish was found to be swimming at $\mbox{St} \approx 0.30$ for $\mbox{Re}$ ranging from $110000$ to $457000$. We use a mesh size $(N_x,N_y,N_z)=(385,129,161)$ for the entire flow domain that translates to $7.99$ million total elements. For the mesh independent study, the readers are referred to the \cite{Liu2017} and \cite{Khalid2020b}.

\section {Results \& Discussion}
\label{sec:rresults}
 In this section, we first present the time-averaged hydrodynamics performance parameters; thrust coefficient ($C_T = {F_T}/{0.5{\rho}{{U_\infty}^2}{A_{CF}}}$), power consumption ($C_P = (\oint{(\boldsymbol{\sigma} \cdot \boldsymbol{n})} \cdot \boldsymbol{V}{ds}) / {0.5{\rho}{{U_\infty}^3}}{A_{CF}}$), and swimming efficiency ($\eta={\overline{C_T}}{U_\infty}/{\overline{}{C_P}}$). The bar over a coefficient represents its respective cycle-averaged value. Next, we discuss the wake topology and vortex dynamics for variations in the wavelength of the prescribed motion of Jack Fish and its real kinematics.

\subsection{Hydrodynamic Performance Parameters} 
\label{subsec:CTCPETA}
\begin{figure}
  \centerline{\includegraphics[scale=0.25]{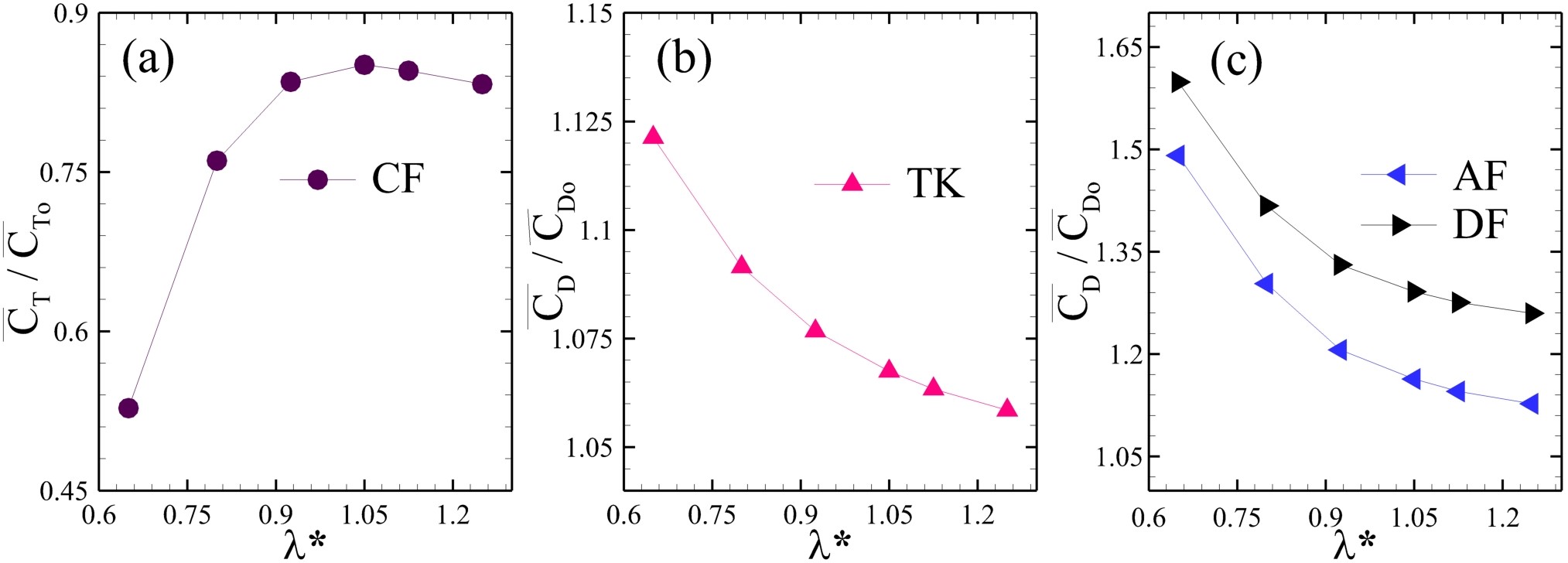}}
  \caption{(a) Thrust coefficient of the caudal fin, (b) drag coefficient of the trunk, and (c) drag coefficient of the anal and dorsal fins, where each cycle-averaged coefficient is normalized by its respective value for the real fish kinematics}
\label{fig:MEANCTCD}
\end{figure}

In our present work, we perform all our simulations for seven undulatory cycles. Hydrodynamic forces on the fish model attain their periodic steady states after three cycles not shown here for brevity. We present the cycle-averaged values of the thrust coefficient ($\overline{C_T}$) for the caudal fin and drag coefficients ($\overline{C_D}$) for the trunk and median fins in Fig.~\ref{fig:MEANCTCD} for Jack Fish with the prescribed motion. These coefficients are normalized by their respective values provided in Table~\ref{tab:real_parameters} for the real kinematics of this fish. It is interesting to note that the caudal fin of Jack Fish produces the highest amount of thrust when it undulates with $\lambda^\star=1.05$. This value of $\lambda^\star$ seems to be an optimal condition and the real fish also undulates with almost the same $\lambda^\star$. As we increase $\lambda^\star$, the drag production by the trunk and median fins reduces. Thus, it can explain why real carangiform fish choose this $\lambda^\star$ in order to swim steadily. 

\begin{table}
  \begin{center}
\def~{\hphantom{0}}
  \begin{tabular}{lcccc}
      Quantity    & CF     &   TK    & AF      & DF \\[3pt]
       $\overline{C_T}$  & 0.2316 &    -    & -       &  - \\
       $\overline{C_D}$  & -      & 0.6373  & 0.0165  & 0.0127\\
       $\overline{C_P}$  & 0.6202 & 0.6401  & 0.0487  & 0.046\\
       $\eta$       & 0.1972 & -0.5257 & -0.1786 & -0.1454 \\
  \end{tabular}
  \caption{Cycle-averaged hydrodynamics performance parameters for the real Jack Fish kinematics}
  \label{tab:real_parameters}
  \end{center}
\end{table}

\begin{figure}
  \centerline{\includegraphics[scale=0.18]{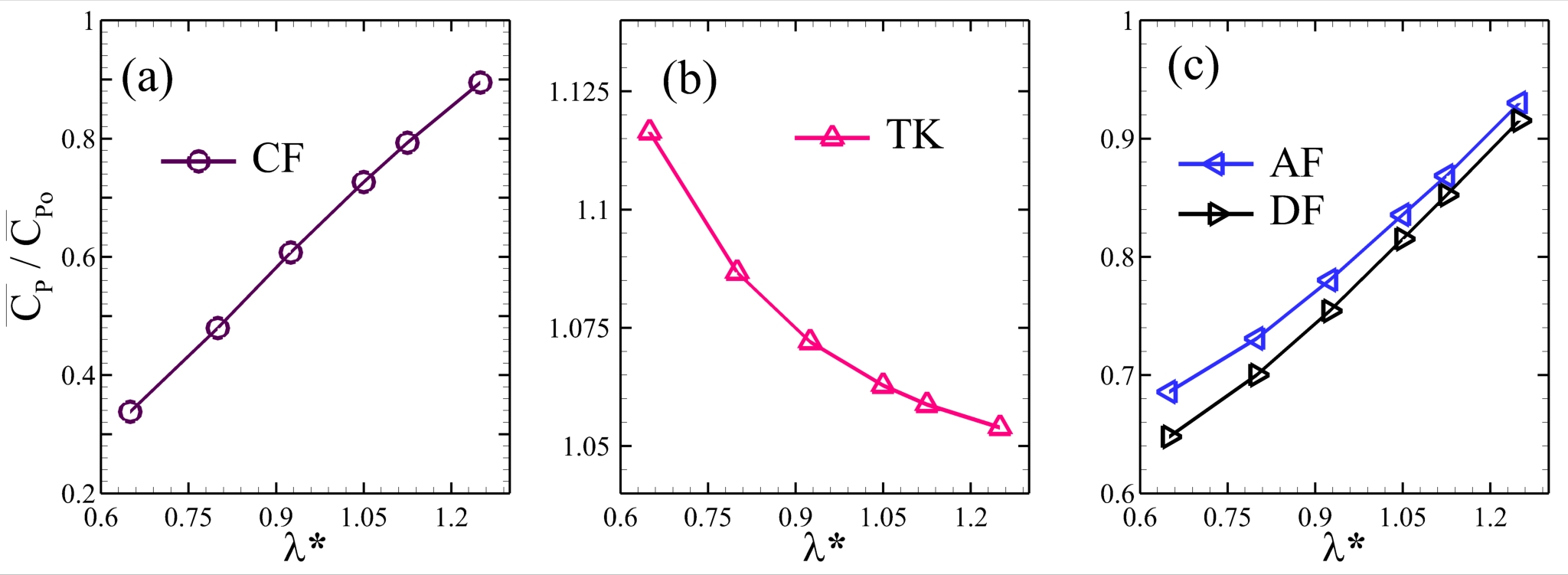}}
  \caption{Cycle-averaged power coefficient of (a) the caudal fin, (b) trunk, and (c) anal and dorsal fins, where each coefficient is normalized by its respective value for the real fish kinematics}
\label{fig:MeanCp}
\end{figure}

Figure~\ref{fig:MeanCp} shows $\overline{C_P}$, normalized by $\overline{C_{P\circ}}$ for all cases with the defined kinematics, where $\overline{C_{P\circ}}$ is the respective power coefficient for each component of the Jack Fish. Greater wavelengths increased power consumption for swimming for all the membranous fins. However, the trunk needs to consume reduced power  when the fish swims with a larger $\lambda^\star$. We also observe that all cases with the prescribed motion show a lower $\overline{C_P}$ than the real kinematics.   

We also show the hydrodynamic efficiency ($\eta$), normalized by $\eta_\circ$, of the caudal fin in Fig.~\ref{fig:EtaCF}. Evidently, the carangiform swimmer is more efficient when swimming with $\lambda^\star=0.80$. Increasing the wavelength would reduce its swimming efficiency. It is interesting that the real swimmer attains the least swimming efficiency compared to those with prescribed kinematics. Thus, efficiency seems to be a less important parameter for steadily swimming fish. It may also be an artifact of how Froude's efficiency is mathematically defined. 

The real kinematics of Jack Fish surpasses the performance of those with the prescribed motion in all cases for thrust production of the caudal fin and drag production for the trunk and median fins. However, individual fins with the real kinematics motion consume more power to achieve the same hydrodynamic motion, while this is reversed for the trunk. Our analysis depicts that natural carangiform swimmers aim to produce more thrust production from their caudal fins while minimizing their efforts to consume less power by their trunks. A wavelength larger than its body-length helps it achieve this superior performance. 
 

\begin{figure}
  \center
  \subfigure[]{\label{fig:EtaCF}\includegraphics[scale=0.08]{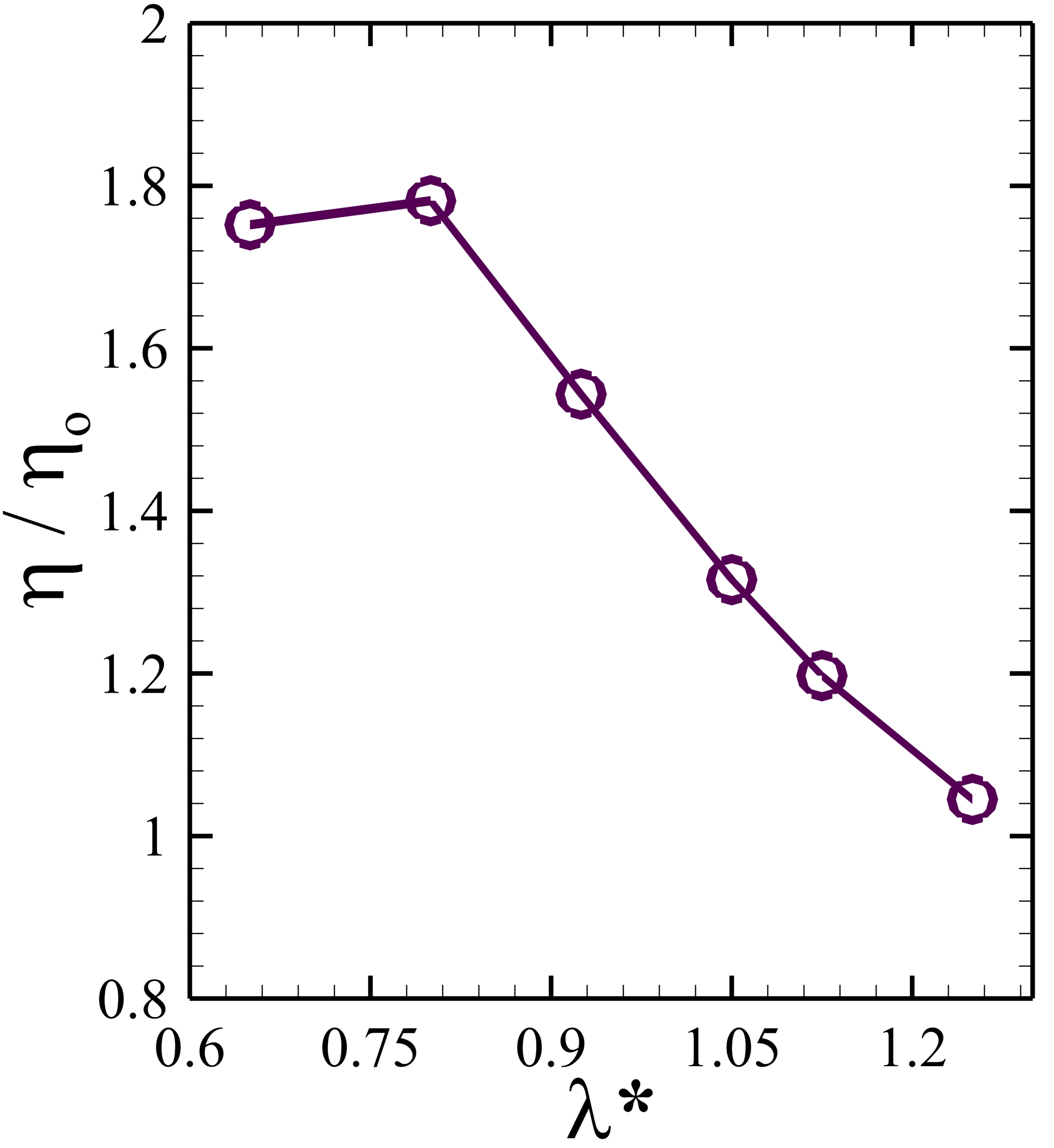}}
  \subfigure[]{\label{fig:temporalCT}\includegraphics[scale=0.1]{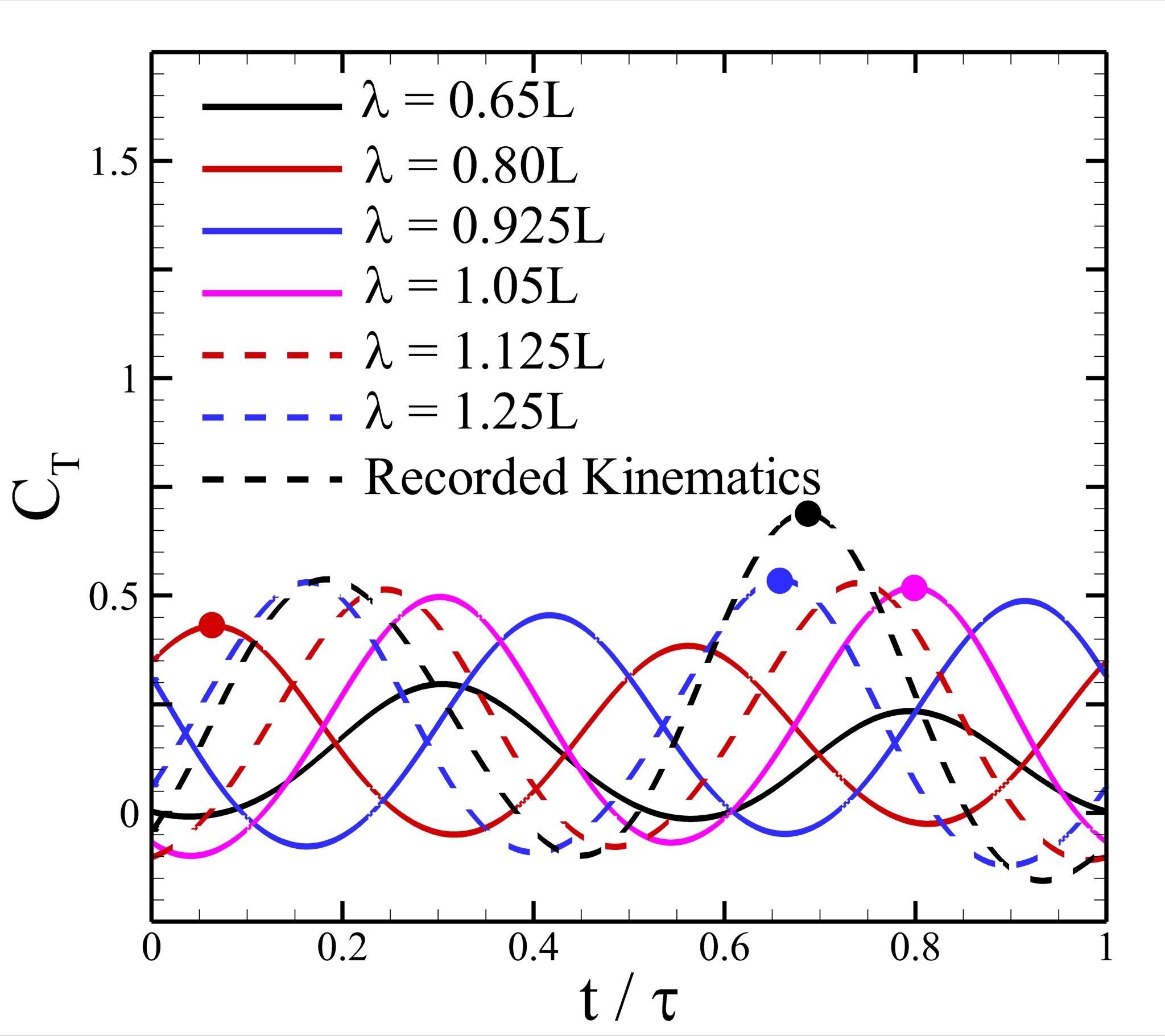}}
 \caption{(a) Hydrodynamic efficiency of the caudal fin of Jack Fish as a function of undulatory wavelength where $\eta$ is normalized by its value for the real fish kinematics and (b) temporal histories of $C_T$ of the caudal fin for the real and prescribed motion}
\label{fig:EtaCF}
\end{figure}

In Fig.~\ref{fig:temporalCT}, we show unsteady thrust of the caudal in one oscillation cycle for all cases. Because the wavelength controls both the oscillation amplitude and phase of the fish body, we see that the peak value of $C_T$ occurs during the first half of the oscillation cycle for $\lambda^\star < 0.925$ and in the second half for greater values of $\lambda^\star$. This requires a more detailed investigation combined with a discussion of the wake dynamics.    

\subsection{Wake Topology and Vortex Dynamics}
\label{subsec:wake}
\begin{figure}
  \centerline{\includegraphics[scale=0.65]{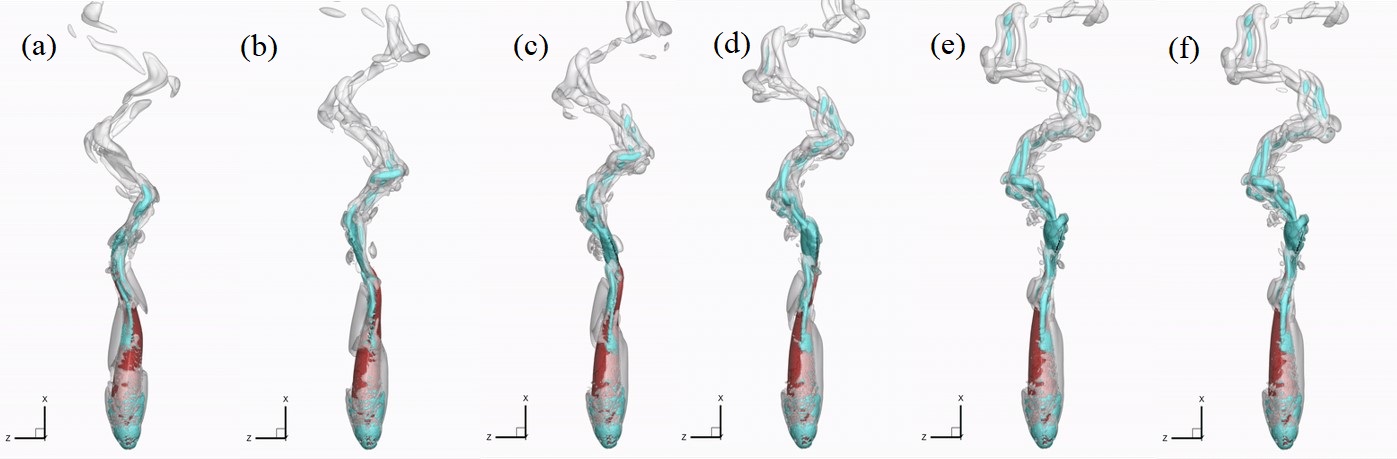}}
 \caption{Top views of three-dimensional vortex structures in the wake of Jack Fish undulating with $\lambda^\star = $ (a) 0.65, (b) 0.80, (c) 0.925, (d) 1.05, (e) 1.125, and (f) 1.25 where the wake structures are colored by the isosurface of Q-criterion. The isosurface $Q = 5$ is in grey and $Q = 35$ is in green. The
latter highlights the vortex core}
\label{fig:wake_comparison_wavelength}
\end{figure}

The physical mechanisms responsible for better thrust production by the caudal fin of Jack Fish swimming with larger $\lambda^\star$ values are explored here using the wake visualizations of the fish swimming.. Figure~\ref{fig:wake_comparison_wavelength} shows coherent vortex structures in the wake of Jack Fish with $\lambda^\star$ varying from $0.65$ to $1.25$. From the top view, it is clear that two interconnected vortex rings are generated and shed in the wake by the swimmer. Similar vortex structures were also observed by \citet{Borazjani2010} for mackarel and by \citet{Han2020} for sunfish, which are carangiform swimmers too. It is evident that the connecting rings get elongated for larger $\lambda^\star$ (see also supplementary movies 1 to 4). Other small vortical structures are also visible in the posterior body regions. 

\begin{figure}
  \centerline{\includegraphics[scale=0.55]{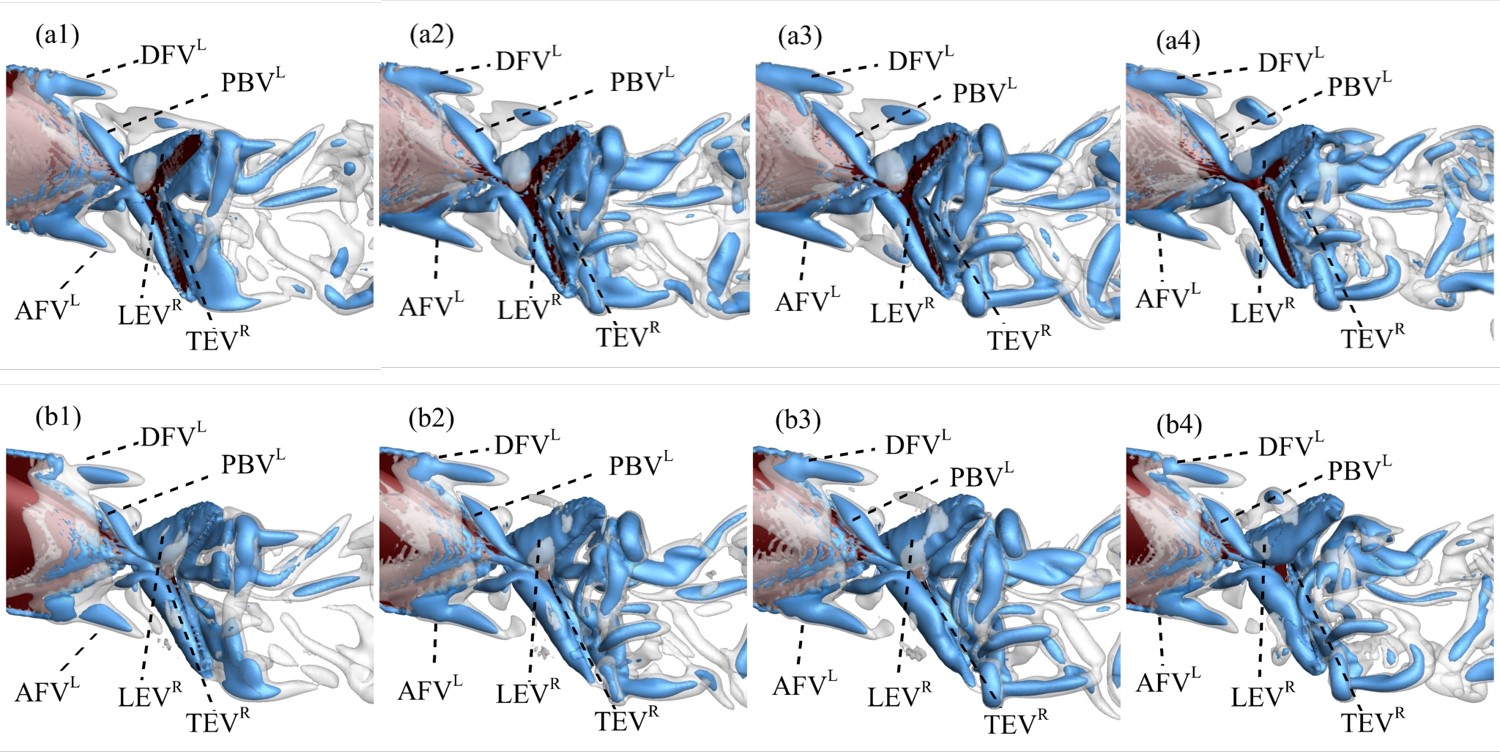}}
 \caption{Vortex dynamics in the vicinity of the caudal fin for Jack Fish undulating with the prescribed motion at $\lambda^\star=0.80$, $1.05$, $1.25$, and the real kinematics in the $1^{\mbox{st}}$, $2^{\mbox{nd}}$, $3^{\mbox{rd}}$, and $4^{\mbox{th}}$ columns, respectively, and the top row shows the flow states at their respective time instants of maximum thrust generation and lower row after a non-dimensional time interval of $0.125$, where the wake structures are colored by the isosurface of Q-criterion. The isosurface $Q = 5$ is in grey and $Q = 35$ is in blue. The latter highlights the vortex core.}
\label{fig:vortex_comparison}
\end{figure}

Figure~\ref{fig:vortex_comparison} focuses on the vortex dynamics in the vicinity of the posterior body and caudal fin for the real kinematics and prescribed motion with $\lambda^\star=0.80$, $1.05$, $1.25$. Here, superscripts ``\textit{L}'' and ``\textit{R}'' represent the vortices produced during the leftward and rightward oscillations, respectively. In each cycle, vortices are shed from the dorsal and anal fins, termed as $\mbox{DFV}$ and $\mbox{AFV}$, respectively, along with the production and traversing of posterior body vortices ($\mbox{PBVs}$). These small-scale vortices are bound to be intercepted by the caudal fin during its oscillations. This capturing of vortices by the caudal fin enhances the strength of the $\mbox{LEV}$s shed from the fin \citep{Liu2017}. Due to the varying oscillation phase for different $\lambda^\star$, we select the timing of the maximum thrust production in all the cases to comparatively analyze the flow dynamics. These time instants are indicated by circles on the plots for unsteady thrust of the caudal fin in Fig.~\ref{fig:temporalCT}. 

The maximum thrust production occurs during initial stages of the rightward flapping stroke for all the kinematic conditions. The top row in Fig.~\ref{fig:vortex_comparison} shows vortex configurations at these time instants. Here, we find that $\mbox{PBVs}$  and $\mbox{LEVs}$ for large $\lambda^\star$ are stronger compared to those for $\lambda^\star=0.80$. The lower row in Fig.~\ref{fig:vortex_comparison} is for flow after $\Delta t /\tau = 0.125$, where the LEVs have grown to cover the whole left side of the caudal fin and subsequently merge with trailing-edge vortices ($\mbox{TEVs}$).

\begin{figure}
  \centerline{\includegraphics[scale=0.55]{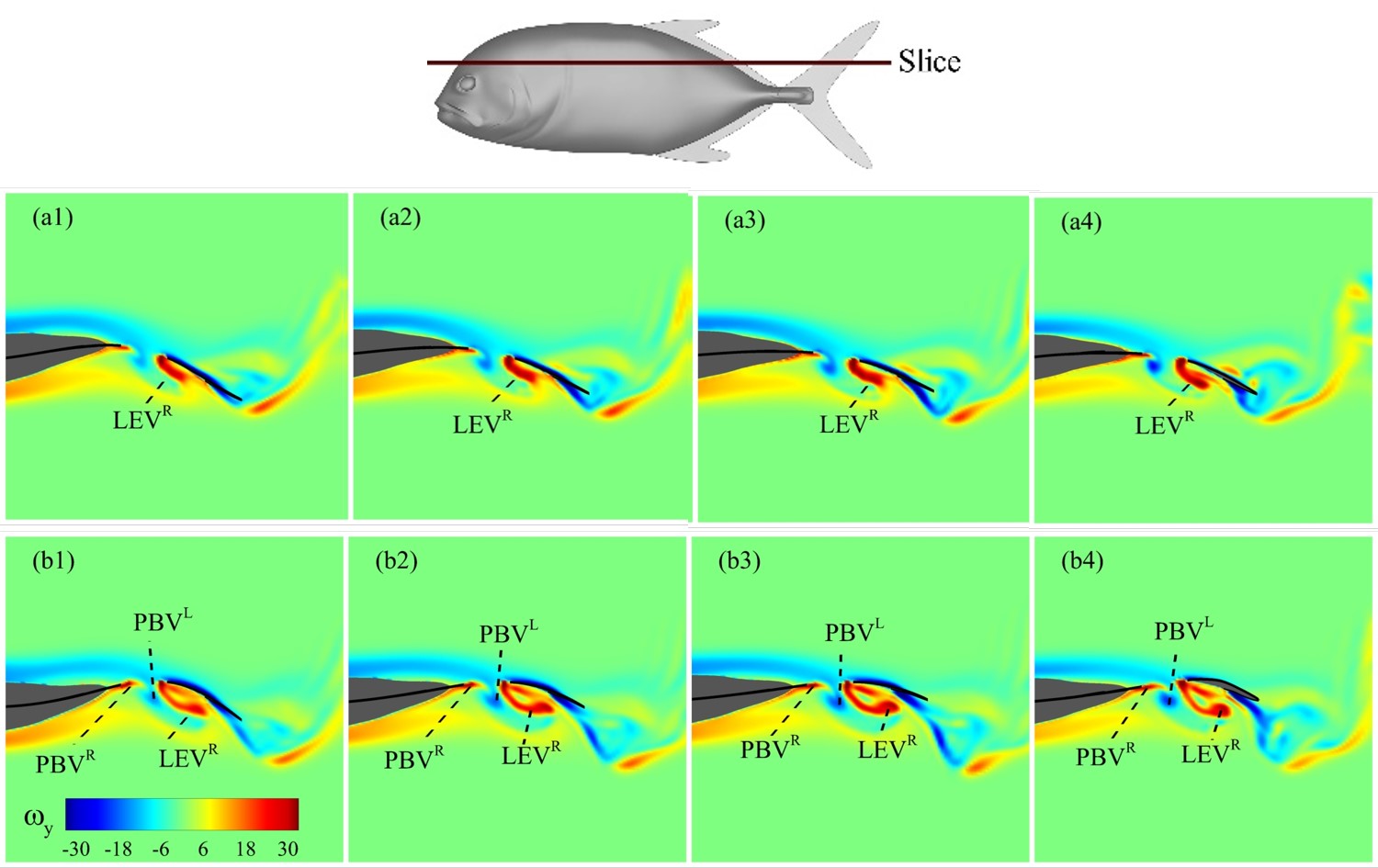}}
 \caption{$\omega_y$ vorticity (computed on the slice as shows by the line on the fish model here) in the vicinity of the caudal fin for Jack Fish undulating with prescribed motion at $\lambda^\star=0.80$, $1.05$, $1.25$, and real kinematics in the $1^{\mbox{st}}$, $2^{\mbox{nd}}$, $3^{\mbox{rd}}$, and $4^{\mbox{th}}$ columns, respectively, and the top row shows the flow states at their respective time instants of maximum thrust generation and lower row after a non-dimensional time interval of $0.125$. }
\label{fig:wyslice}
\end{figure}

\begin{figure}
  \centerline{\includegraphics[scale=0.15]{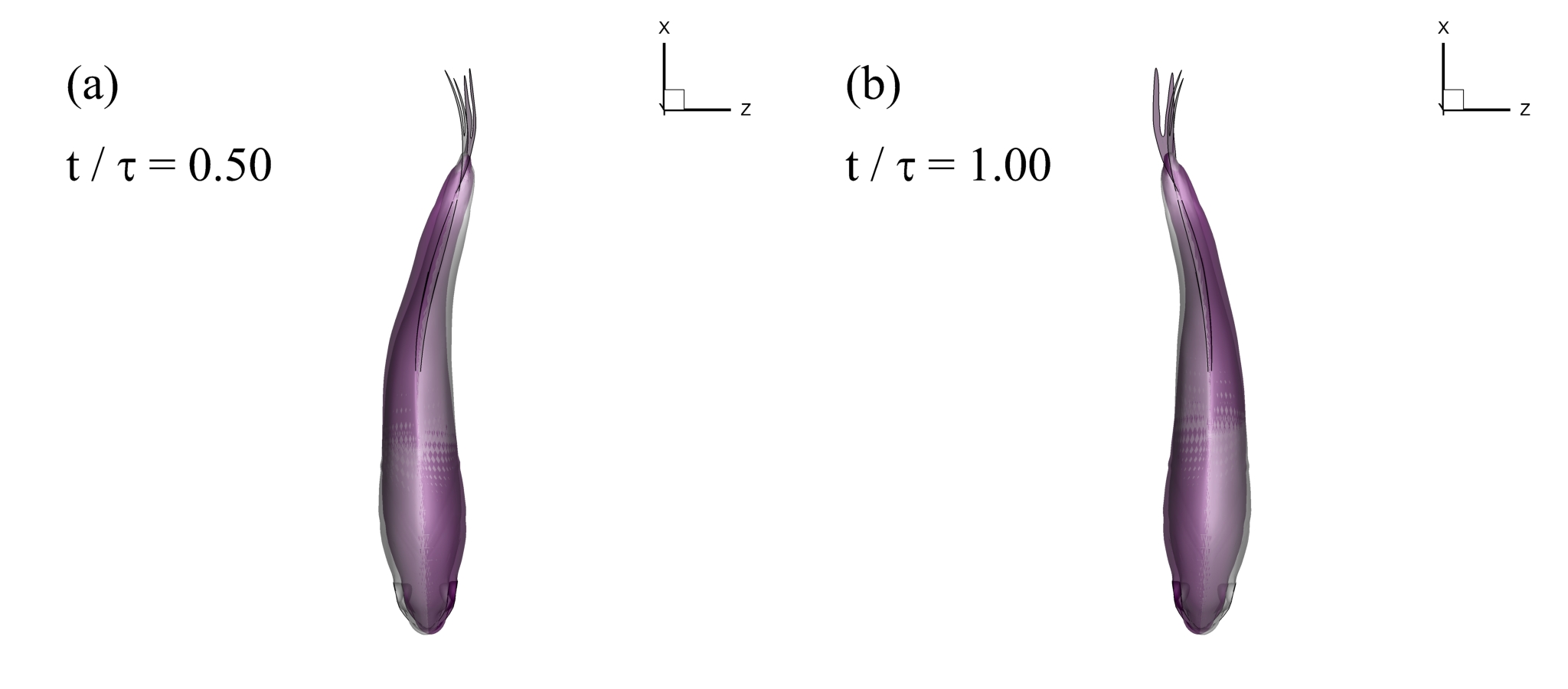}}
 \caption{Comparison of the real and prescribed ($\lambda^\star=1.05$) motion of Jack Fish (with models in dark and light colors, respectively) at two time instants} 
\label{fig:comaprison_wavy}
\end{figure}

Next, we quantitatively examine the y-component of vorticity ($\omega_y$) by extracting a slice at the location shown in Fig.~\ref{fig:wyslice} at the same time instants as in Fig.~\ref{fig:vortex_comparison}. The contours show the elevated LEVs' strength for larger $\lambda^\star$ values. This figure also provides visualizations for the interference of LEVs at the caudal fin with PBVs produced during the previous stroke. To further quantify the strength of $\mbox{LEVs}$, we compute their circulation ($\Gamma$) using contours of $\omega_y$ during rightward flapping strokes using a technique for the identification of vortices explained by \cite{Khalid2020a} and plot in Fig.~\ref{fig:wyslice}. We observe that larger wavelengths produce $\mbox{LEVs}$ with more circulation (see also supplementary movies 5 to 8). During this stroke, posterior parts of LEVs are detached from their main cores attached to the leading-edge of the caudal fin. Main cores of $\mbox{LEVs}$ recover their strength in the later stages of the stroke. Out of all the cases with the prescribed motion, $\lambda^\star=1.05$ shows maximum $\Gamma$. It elaborates higher thrust production under these conditions. Owing to the flexibility of the caudal fin \citep{Liu2017}, the $\mbox{LEV}$ is the strongest for Jack Fish with the real kinematics. Figure~\ref{fig:comaprison_wavy} exhibits the comparison of forms of Jack Fish with the real and prescribed motion at $\lambda^\star = 1.05$ for two time instants. We notice that the dorsal and ventral sides of the caudal fin show asymmetry in their oscillation amplitudes and it could happen due to the felxible membranous structure of the fish's tail. It is consistent with recent reports where flexible structures could produce higher thrust as compared to their rigid counterparts \citep{White2020}. 

\begin{figure}
  \centerline{\includegraphics[scale=0.1]{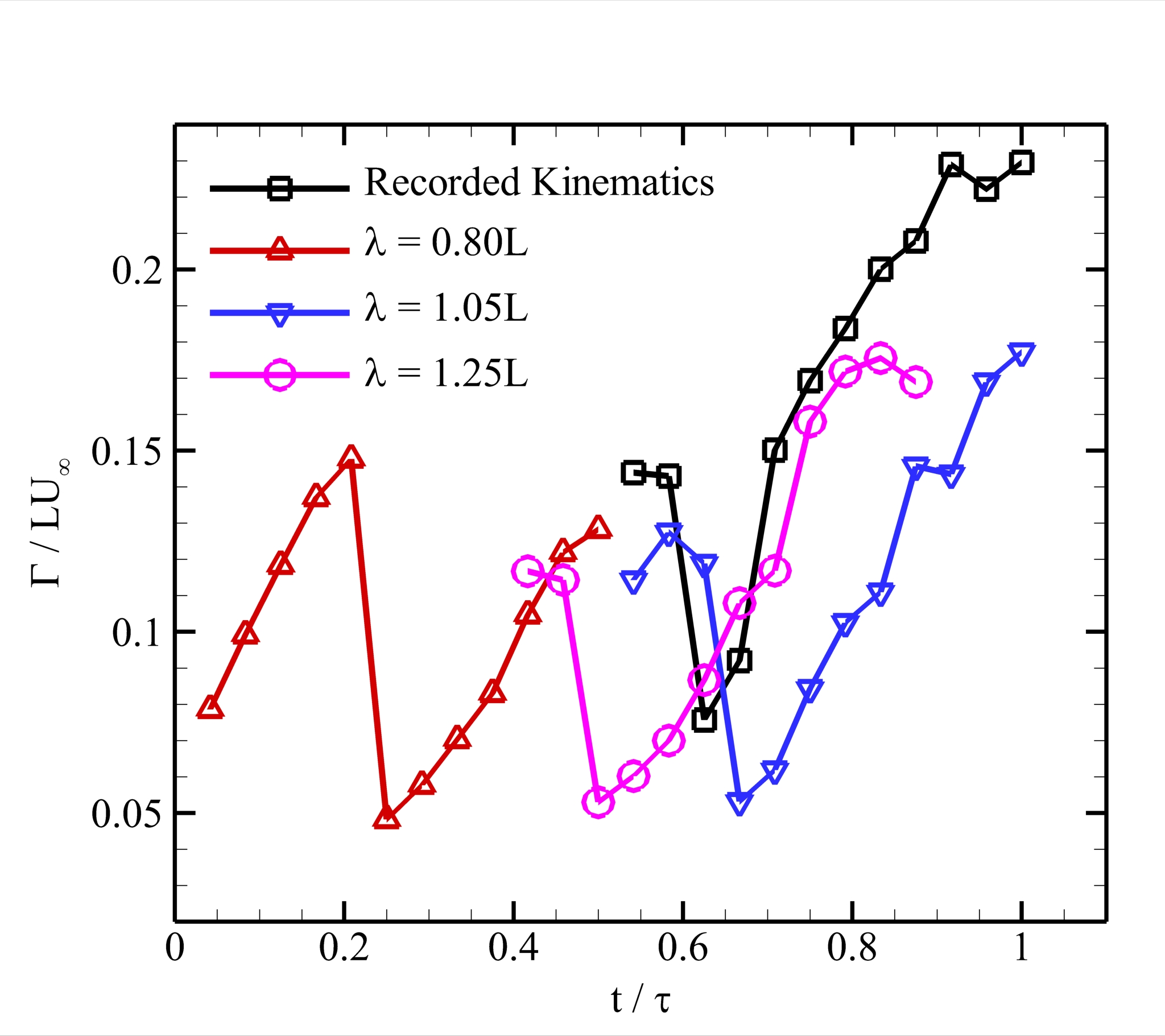}}
 \caption{Circulation of $\mbox{LEVs}$ that remain attached with the caudal fin throughout its rightward flapping stroke}
\label{fig:kd}
\end{figure}

\section{Conclusions}
In order to study hydrodynamics of Jack Fish, we reconstructed its physiological model and kinematics. We further utilize this model to prescribe carangiform fish-like kinematics and perform high-fidelity numerical simulations for flows over these models for a range of wavelengths. The results show that the carangiform swimmer chooses to swim with $\lambda^\star \approx 1.05$ to (1) produce more thrust production by its caudal fin, (2) alleviate the drag generation for its trunk, and (3) minimize the consumption of power from its trunk. This study also reveals that the real swimmer has lower swimming efficiency compared its counterparts with the prescribed kinematics. Our qualitative and quantitative analyses also demonstrate that stronger vortices are produced by this swimmer when it undulates with $\lambda > L$. Nevertheless, $\lambda \approx 1.05$ provides an optimal condition for this swimmer to fulfill the aforementioned objectives. We also notice that the fish with its real kinematics outperforms others due to the flexiblity of the caudal fin that helps $\mbox{LEVs}$ remain attached to their leading-edge, which enables them to recover quickly even when their posterior part detaches and is shed in the wake. 

\backsection[Acknowledgements]{M. S. U. Khalid is International Exchange Postdoctoral Research Fellow sponsored by China National Science Postdoc Foundation and Peking University. H. Dong acknowledges the support from NSF CNS grant no. CPS-1931929 and SEAS Research Innovation Awards of the University of Virginia.}


\backsection[Declaration of interests]{The authors report no conflict of interest.} 



\bibliographystyle{jfm}
\bibliography{jfm}

\begin{thebibliography}{15}
\expandafter\ifx\csname natexlab\endcsname\relax\def\natexlab#1{#1}\fi
\def\au#1{#1} \def\ed#1{#1} \def\yr#1{#1}\def\at#1{#1}\def\jt#1{\textit{#1}}
  \def\bt#1{#1}\def\bvol#1{\textbf{#1}} \def\vol#1{#1} \def\pg#1{#1}
  \def\publ#1{#1}\def\arxiv#1{#1}\def\org#1{#1}\def\st#1{\textit{#1}}

\bibitem[Borazjani \& Sotiropoulos(2010)]{Borazjani2010}
{\sc \au{Borazjani, I.} \& \au{Sotiropoulos, F.}} \yr{2010}  \at{On the role of
  form and kinematics of on the hydrodynamics of self-propelled body/caudal fin
  swimming}.  \jt{Journal of Experimental Biology}  \bvol{213}~(1),
  \pg{89--107}.

\bibitem[Fish(2020)]{Fish2020}
{\sc \au{Fish, F.}} \yr{2020}  \at{Advantages of aquatic animals as models for
  bio-inspired drones over present auv technology}.  \jt{Bioinspiration \&
  BIomimetics}  \bvol{15}~(2),  \pg{025001}.

\bibitem[Han {\em et~al.\/}(2020)Han, Lauder \& Dong]{Han2020}
{\sc \au{Han, P.}, \au{Lauder, G.~V.} \& \au{Dong, H.}} \yr{2020}
  \at{Hydrodynamics of median-fin interactions in fish-like locomotion: Effects
  of fin shape and movement}.  \jt{Physics of Fluids}  \bvol{32}~(1),
  \pg{011902}.

\bibitem[Khalid {\em et~al.\/}(2016)Khalid, Akhtar \& Dong]{Khalid2016}
{\sc \au{Khalid, M. S.~U.}, \au{Akhtar, I.} \& \au{Dong, H.}} \yr{2016}
  \at{Hydrodynamics of a tandem fish school with asynchronous undulation of
  individuals}.  \jt{Journal of Fluids and Structures}  \bvol{66},
  \pg{19--35}.

\bibitem[Khalid {\em et~al.\/}(2018)Khalid, Akhtar, Imtiaz, Dong \&
  Wu]{Khalid2018}
{\sc \au{Khalid, M. S.~U.}, \au{Akhtar, I.}, \au{Imtiaz, H.}, \au{Dong, H.} \&
  \au{Wu, B.}} \yr{2018}  \at{On the hydrodynamics and nonlinear interaction
  between fish in tandem configuration}.  \jt{Ocean Engineering}  \bvol{157},
  \pg{108--120}.

\bibitem[Khalid {\em et~al.\/}(2020{\natexlab{{\em a\/}}})Khalid, Wang, Akhtar,
  Dong \& Liu]{Khalid2020b}
{\sc \au{Khalid, M. S.~U.}, \au{Wang, J.}, \au{Akhtar, I.}, \au{Dong, H.} \&
  \au{Liu, M.~B.}} \yr{2020{\natexlab{{\em a\/}}}}  \at{Modal decompositions of
  the kinematics of crevalle jack and the fluid-caudal fin interaction}.
  \jt{Bioinspiration \& Biomimetics} .

\bibitem[Khalid {\em et~al.\/}(2020{\natexlab{{\em b\/}}})Khalid, Wang, Dong \&
  Liu]{Khalid2020a}
{\sc \au{Khalid, M. S.~U.}, \au{Wang, J.}, \au{Dong, H.} \& \au{Liu, M.~B.}}
  \yr{2020{\natexlab{{\em b\/}}}}  \at{Flow transitions and mapping for
  undulating swimmers}.  \jt{Physical Review Fluids}  \bvol{5}~(6),
  \pg{063104}.

\bibitem[Lauder \& Madden(2006)]{Lauder2006a}
{\sc \au{Lauder, G.~V.} \& \au{Madden, P. G.~A.}} \yr{2006}  \at{Learning from
  fish: Kinematics and experimental hydrodynamics for robotics}.
  \jt{International Journal of Automation and Computing}  \bvol{4},
  \pg{325--335}.

\bibitem[Liu {\em et~al.\/}(2017)Liu, Ren, Dong, Akanyati, Liao \&
  Lauder]{Liu2017}
{\sc \au{Liu, G.}, \au{Ren, Y.}, \au{Dong, H.}, \au{Akanyati, O.}, \au{Liao,
  J.~C.} \& \au{Lauder, G.~V.}} \yr{2017}  \at{Computational analysis of vortex
  dynamics and performance enhancement due to body-fin and fin-fin interactions
  in fish-like locomotion}.  \jt{J. Fluid. Mech.}  \bvol{829},  \pg{65--88}.

\bibitem[Mittal {\em et~al.\/}(2008)Mittal, Dong, Bozkurttas, Najjar, Vargas \&
  Loebbecke]{Mittal2008}
{\sc \au{Mittal, R.}, \au{Dong, H.}, \au{Bozkurttas, M.}, \au{Najjar, F.~M.},
  \au{Vargas, A.} \& \au{Loebbecke, A.~Von}} \yr{2008}  \at{A versatile sharp
  interface immersed boundary method for incompressible flows with complex
  boundaries}.  \jt{Journal of Computational Physics}  \bvol{227}~(10),
  \pg{4825--4852}.

\bibitem[Sfakiotakis {\em et~al.\/}(1999)Sfakiotakis, Lane \&
  Davies]{Sfakiotakis1999}
{\sc \au{Sfakiotakis, M.}, \au{Lane, D.~M.} \& \au{Davies, J. B.~C.}} \yr{1999}
   \at{Review of fish swimming modes for aquatic locomotion}.  \jt{IEEE JOURNAL
  OF OCEANIC ENGINEERING}  \bvol{24}~(2),  \pg{237--252}.

\bibitem[Videler(1993)]{Videler1993}
{\sc \au{Videler, J.~J.}} \yr{1993} {\em Fish swimming\/}.  \publ{Chapman and
  Hall, London}.

\bibitem[Wang {\em et~al.\/}(2020)Wang, Ren, Li \& Dong]{Wang2020}
{\sc \au{Wang, J.}, \au{Ren, Y.}, \au{Li, C.} \& \au{Dong, H.}} \yr{2020}
  \at{Tuna locomotion: a computational hydrodynamic analysis of finlet
  function}.  \jt{Journal of the Royal Society Interface}  \bvol{17}~(165),
  \pg{20190590}.

\bibitem[White {\em et~al.\/}(2020)White, , Lauder \& Bart-Smith]{White2020}
{\sc \au{White, C.}, , \au{Lauder, G.~V.} \& \au{Bart-Smith, H.}} \yr{2020}
  \at{Tunabot flex: a tuna-inspired robot with body flexibility improves
  high-performance swimming}.  \jt{Bioinspiration \& Biomimetics} .

\bibitem[Zhu {\em et~al.\/}(2019)Zhu, White, Wainwright, Santo, Lauder \&
  Bart-Smith]{Zhu2019}
{\sc \au{Zhu, J.}, \au{White, C.}, \au{Wainwright, D.~K.}, \au{Santo, V.~Di},
  \au{Lauder, G.~V.} \& \au{Bart-Smith, H.}} \yr{2019}  \at{Tuna robotics: A
  high-frequency experimental platform exploring the performance space of
  swimming fishes}.  \jt{Science Robotics}  \bvol{4}~(34),  \pg{4615}.

\end{thebibliography}

\end{document}